\documentclass[english,aps,prl,nofootinbib,twocolumn,citeautoscript,reprint]{revtex4-1}

\setcitestyle{super}

\usepackage[latin9]{inputenc}
\usepackage{verbatim}
\usepackage{graphicx}
\usepackage{xargs}[2008/03/08]
\usepackage{xcolor}



\usepackage{babel}

\begin{document}

\makeatletter 
\renewcommand\@biblabel[1]{#1.}
\makeatother

\newcommand*{\citen}[1]{%
  \begingroup
    \romannumeral-`\x 
    \setcitestyle{numbers}%
    \citealp{#1}%
  \endgroup   
}

\newcommandx\ket[1][usedefault, addprefix=\global, 1=]{|#1\rangle}
\newcommandx\bra[1][usedefault, addprefix=\global, 1=]{\langle#1|}
\newcommandx\avg[1][usedefault, addprefix=\global, 1=]{\langle#1\rangle}
\newcommandx\var[1][usedefault, addprefix=\global, 1=]{\langle(\Delta#1)^{2}\rangle}
\global\long\def\aa{\hat{a}}
\global\long\def\ad{\hat{a}^{\dagger}}

\title{Bandwidth manipulation of quantum light by an electro-optic time lens}

\author{Micha\l{} Karpi\'{n}ski{$^{1,2,*}$}}
\author{Micha\l{} Jachura{$^{1,2,*}$}}
\author{Laura J. Wright{$^{1}$}}
\author{Brian J. Smith{$^{1,3}$}}

\affiliation{{$^{1}$}Clarendon Laboratory, University of Oxford, Parks Road, Oxford, OX1 3PU, UK}
\affiliation{{$^{2}$}Faculty of Physics, University of Warsaw, Pasteura 5, 02-093 Warszawa, Poland}
\affiliation{{$^{3}$}Department of Physics and Oregon Center for Optical, Molecular, and Quantum Science, University of Oregon, Eugene, Oregon 97403, USA}
\affiliation{{$^{*}$} These authors contributed equally to this work.}

\maketitle

\textbf{The ability to manipulate the spectral-temporal waveform of optical pulses has enabled a wide range of applications from ultrafast spectroscopy\cite{Mukamel2013} to high-speed communications\cite{Weiner2011}. Extending these concepts to quantum light has the potential to enable breakthroughs in optical quantum science and technology\cite{Kielpinski2011,Roslund2013,Brecht2015}. However, filtering and amplifying often employed in classical pulse shaping techniques are incompatible with non-classical light. Controlling the pulsed mode structure of quantum light requires efficient means to achieve deterministic, unitary manipulation that preserves fragile quantum coherences. Here we demonstrate an electro-optic method for modifying the spectrum of non-classical light by employing a time lens\cite{Kolner1994, Foster2009, Torres-Company2011}. In particular we show highly-efficient wavelength-preserving six-fold compression of single-photon spectral intensity bandwidth, enabling over a two-fold increase of single-photon flux into a spectrally narrowband absorber. These results pave the way towards spectral-temporal photonic quantum information processing and facilitate interfacing of different physical platforms\cite{Kimble2008,Duan2001,Lavoie2013} where quantum information can be stored\cite{Lvovsky2009} or manipulated\cite{Campbell2014}.
}

The time-frequency (TF) degree of freedom of non-classical light has come to the fore as a promising candidate for multidimensional quantum information science\cite{Kielpinski2011,Roslund2013,Brecht2015}, due in part to its compatibility with integrated photonic platforms and fibre networks\cite{Edition}.
 Recent research efforts have diversified from generation of single-photon wavepackets with controlled TF properties\cite{Mosley2008} and their characterization\cite{Wasilewski2007} to active modification of the TF state of quantum light. Tremendous progress has been made in techniques to shift the central frequency of quantum light, using nonlinear optical methods\cite{Kumar1992, McGuinness2010, Zaske2012, Ates2012, Lavoie2013, Vollmer2014, Matsuda2016}. Going beyond frequency conversion towards pulse shaping, for example bandwidth or general spectral-amplitude manipulation, has proven challenging for nonlinear optical methods to realize with low-noise, deterministic, broad spectral range operation required for non-classical light.

Essential to nearly all optical experiments is the concept of mode matching, whether to achieve high-visibility interference or strong absorption. In the TF domain, spectral mode matching must be achieved for efficient interfacing between physical systems whose optical emission varies both in central wavelength and characteristic bandwidths, such as quantum dots and atomic vapours. Spectral bandwidth manipulation is an essential capability for interfacing different systems with characteristic bandwidths ranging from MHz to THz\cite{Jensen2011,Bustard2013}. Here we experimentally demonstrate TF manipulation of heralded single-photon wavepackets in a low-loss, all-fibre electro-optic platform, which is intrinsically free from optical noise and does not spectrally shift the central wavelength of the pulse. We apply an electro-optic time lens\cite{Kolner1994, Torres-Company2011} to single-photon pulses, achieving wavelength-preserving six-fold bandwidth compression of single-photon states in the near-infrared spectral region.

The deterministic nature of the electro-optic effect, in which temporal phase is applied to each optical pulse, implies unit internal conversion efficiency of this approach, with the overall efficiency limited by transmission losses (Supplementary Information), providing a realistic path to overall unitary operation. This allows us to demonstrate a two-fold increase of photon flux into a spectrally narrowband absorber, confirming practical value of the scheme for the development of quantum networks\cite{Kimble2008}, where losses can be tolerated in a repeat-until-success approach. 

\begin{figure*}[t]
\includegraphics[width=1.3\columnwidth]{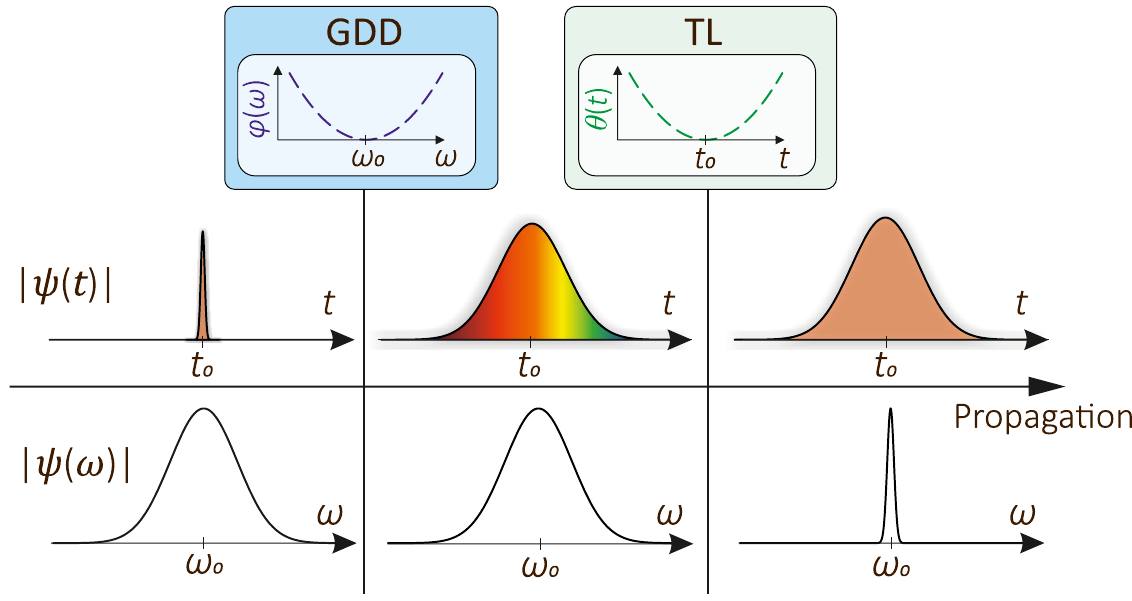}
\caption{\textbf{Conceptual scheme of electro-optic bandwidth compression.} 
An optical wavepacket propagates through a second-order dispersive medium imposing quadratic spectral phase, known as group delay dispersion (GDD). This extends the temporal duration of the single-photon wavepacket $\psi(t)$ by shifting different frequency components in time by a value proportional to the derivative of the spectral phase imprint $\varphi(\omega)$. In the next step a quadratic time-dependent phase profile $\theta(t)$, known as a time lens (TL), is applied to the wavepacket in a synchronous manner shifting the frequency components by a value proportional to its derivative towards the central value $\omega_{0}$, thus compressing the spectral bandwidth of the wavepacket. The amount of GDD is precisely adjusted to fulfill the bandwidth compression condition (see text for details).}
\end{figure*}  

Electro-optic bandwidth compression can be explained by referring to the fundamental space-time duality between paraxial propagation of an optical beam and dispersive propagation of an optical pulse\cite{Kolner1994, Torres-Company2011}. Compression of spatial bandwidth of a Gaussian beam can be accomplished by collimating it using a lens placed one focal length from its waist. This corresponds to multiplication of the transverse field profile by a phase factor quadratic in transverse momentum, acquired during free-space propagation towards the lens, followed by a phase factor quadratic in transverse position, introduced by the lens.
In the space-time duality free-space propagation corresponds to pulse propagation in a second-order dispersive medium imposing a quadratic spectral phase $\varphi(\omega)=\Phi(\omega-\omega_{0})^{2}/2$, where $\Phi$ is the group delay dispersion (GDD) and $\omega_{0}$ the central frequency of the pulse. The
action of a time lens (TL) is realized by quadratic temporal phase centred on the pulse reference frame
$\theta(t)=K t^{2}/2$, where $K$ is a constant chirping factor and $t$ the time in the moving pulse reference frame.
Such a phase profile can be realized either via nonlinear interaction with an intense auxiliary pulse\cite{Lavoie2013,Matsuda2016,Foster2009} or electro-optic phase modulation\cite{Kolner1994,Torres-Company2011}, which is used in our experiment. In the TF domain dispersive propagation distributes spectral components of the wavepacket in time, whereas the TL introduces appropriate time-dependent frequency shears that move spectral components towards the central frequency leading to bandwidth compression, as conceptually depicted in Fig.~1. Thus dispersive propagation followed by a TL achieves spectral bandwidth compression of a Gaussian wavepacket, provided the collimation condition $\Phi=1/K$ is satisfied.

\begin{figure*}
\includegraphics[width=1.2\columnwidth]{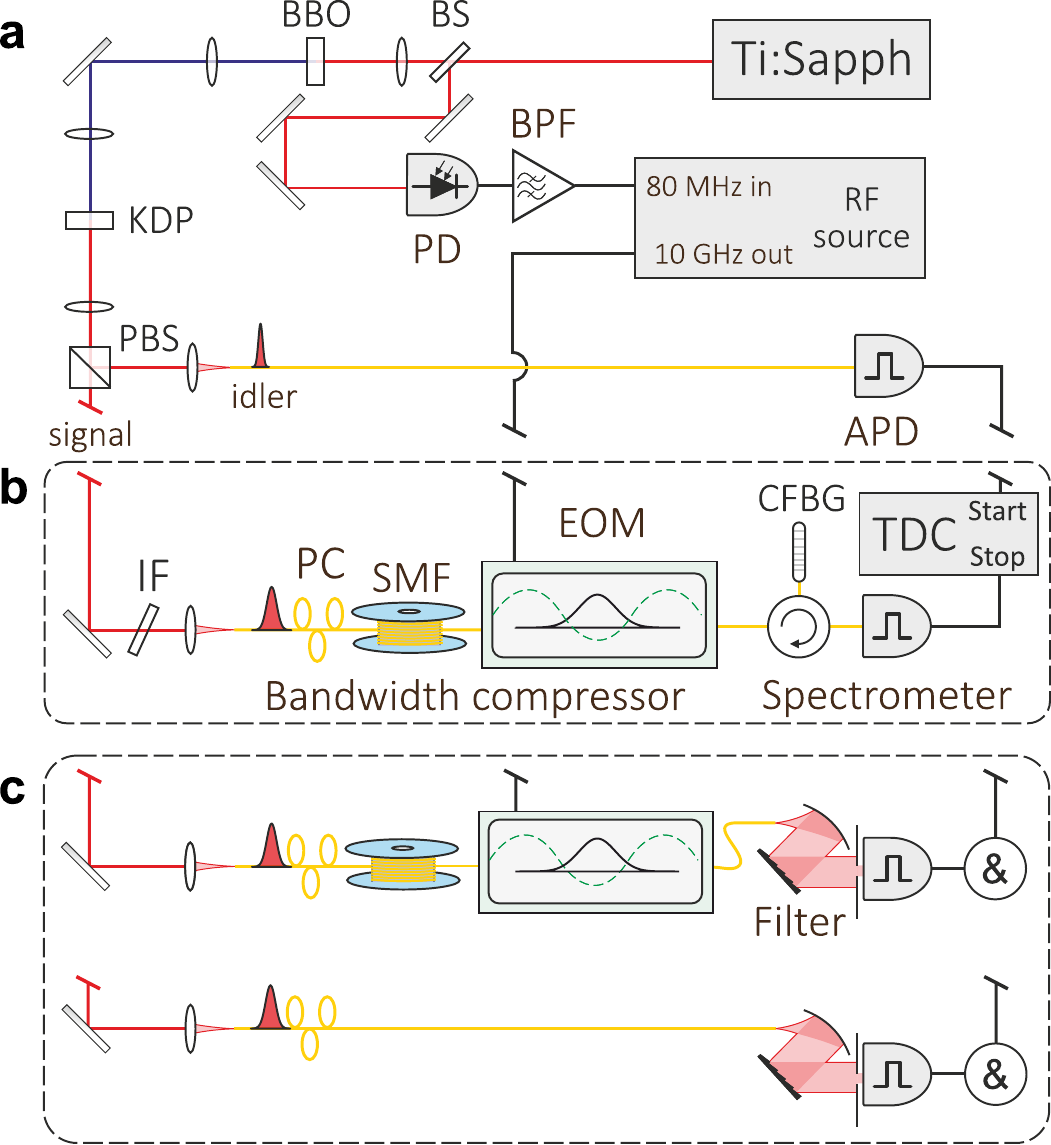}
\caption{\textbf{Experimental setup.} \textbf{a,} Pulses from a Ti:Sapphire laser frequency doubled in a beta barium borate (BBO) crystal pump SPDC in potassium dihydrogen phosphate (KDP) crystal to generate spectrally-pure single-photon wavepackets in the signal mode, which are heralded by detection of a photon in the idler mode at an avalanche photodiode (APD). \textbf{b,} The signal photons, after separation from the idler photons at a polarizing beam-splitter (PBS) and passing through a bandpass interference filter (IF), are directed towards the EOBC consisting of a polarization controller (PC), single-mode fibre (SMF) and electro-optic phase modulator (EOM). The EOM is driven by a $10$~GHz sinusoidal RF field temporally-locked to the single-photon pulse train, which is derived from a photodiode (PD) monitoring the laser pulse train. Synchronizing the parabolic region of the RF modulation with the single-photon pulse implements a time lens. The initial and final spectra of the heralded single-photon wavepackets are measured by means of a time-of-flight spectrometer comprised of a circulator, chirped fibre Bragg grating (CFBG), low-timing-jitter APD and time-to-digital converter (TDC) working in the start-stop mode. \textbf{c,} To verify the efficiency of the device we compare the total flux of photons through a narrowband spectral filter based on a grating monochromator with and without the use of the EOBC. }
\end{figure*}

To demonstrate this approach to bandwidth manipulation for quantum pulses, we generate fibre-coupled spectrally-pure single-photon wavepackets of $3$~nm full-width at half-maximum (FWHM) spectral bandwidth and $830$~nm central wavelength by heralding from a pulsed spontaneous parametric down-conversion (SPDC) single-photon source\cite{Mosley2008} (Methods). The electro-optic bandwidth compressor (EOBC) is realized by propagating the heralded photon through $256$~m of standard single-mode fibre introducing $9.9\mathrm{\:ps^{2}}$ GDD followed by a lithium-niobate-waveguide traveling-wave electro-optic phase modulator (EOM), Fig.~2b. The TL is realized by driving the modulator
with a $f_{\mathrm{RF}}=10\mathrm{\:GHz}$, $33$~dBm sinusoidal radio-frequency (RF) signal, yielding temporal phase modulation
of the form $\theta_{\mathrm{RF}}(t)=A\sin(2\pi f_{\mathrm{RF}}t)$,
where $A=25.7$~rad (Supplementary Information). The RF signal is produced by amplifying the output
of a dielectric resonator oscillator phase-locked to a high harmonic
of an $80$~MHz sinusoidal reference signal derived from the pulsed SPDC pump beam\cite{Kang2003}. The relative temporal position of the optical pulse and RF signal was adjusted by an RF phase shifter. As the duration
of the optical pulse is shorter than the modulation
period, the TL is approximated by the parabolic region of sinusoidal phase modulation, known as the TL aperture\cite{Kolner1994}, giving a chirping factor of $K=4\pi^{2}f_{\mathrm{RF}}^{2}A$, see Fig.~2. To just fill the TL aperture after propagation through the dispersive medium the bandwidth of heralded photons was reduced to $0.9$~nm FWHM using an interference filter (IF). The GDD value was chosen to match the TL chirping factor (Supplementary Information).

To directly verify this approach to bandwidth manipulation the spectrum of heralded single-photon wavepackets was measured before and after the compression procedure using an efficient $0.07$~nm resolution spectrometer\cite{Davis2016} (Fig.~2 and  Methods). The spectrum of heralded photons entering the EOBC was acquired at the EOM output with no RF signal applied. Subsequently the RF signal was switched on and the phase set to implement a focusing TL, followed by heralded single photon spectrum acquisition. To further illustrate the space-time duality between electro-optic bandwidth compression and Gaussian beam collimation the heralded single-photon spectrum for a diverging TL was obtained. Measured spectra for the three aforementioned cases (initial, focusing TL and diverging TL), yielding FWHM bandwidths of $0.92\pm0.06$~nm ($401\pm26$~GHz), $0.15\pm0.01$~nm ($65\pm4$~GHz) and $1.45\pm0.17$~nm ($631\pm74$~GHz), respectively, are presented in Fig.~3a (see Methods). We quantify the device performance using a compression factor given by the input and output spectral bandwidth ratio. Due to finite spectral resolution the $6.1\pm0.6$ ratio extracted from raw measured data presents a lower bound on the EOBC performance. The experimentally determined spectrometer instrument response function yields a FWHM resolution of $0.07$~nm, Fig.~3b. 

By modifying the TL chirping rate and aperture as well as the GDD the EOBC can be tuned to a different bandwidth regime. In Supplementary Information we show spectral compression results of $2$-nm bandwidth pulses by means of a $40$-GHz electro-optic TL. 

\begin{figure}[!th]
\includegraphics[width=0.88\columnwidth]{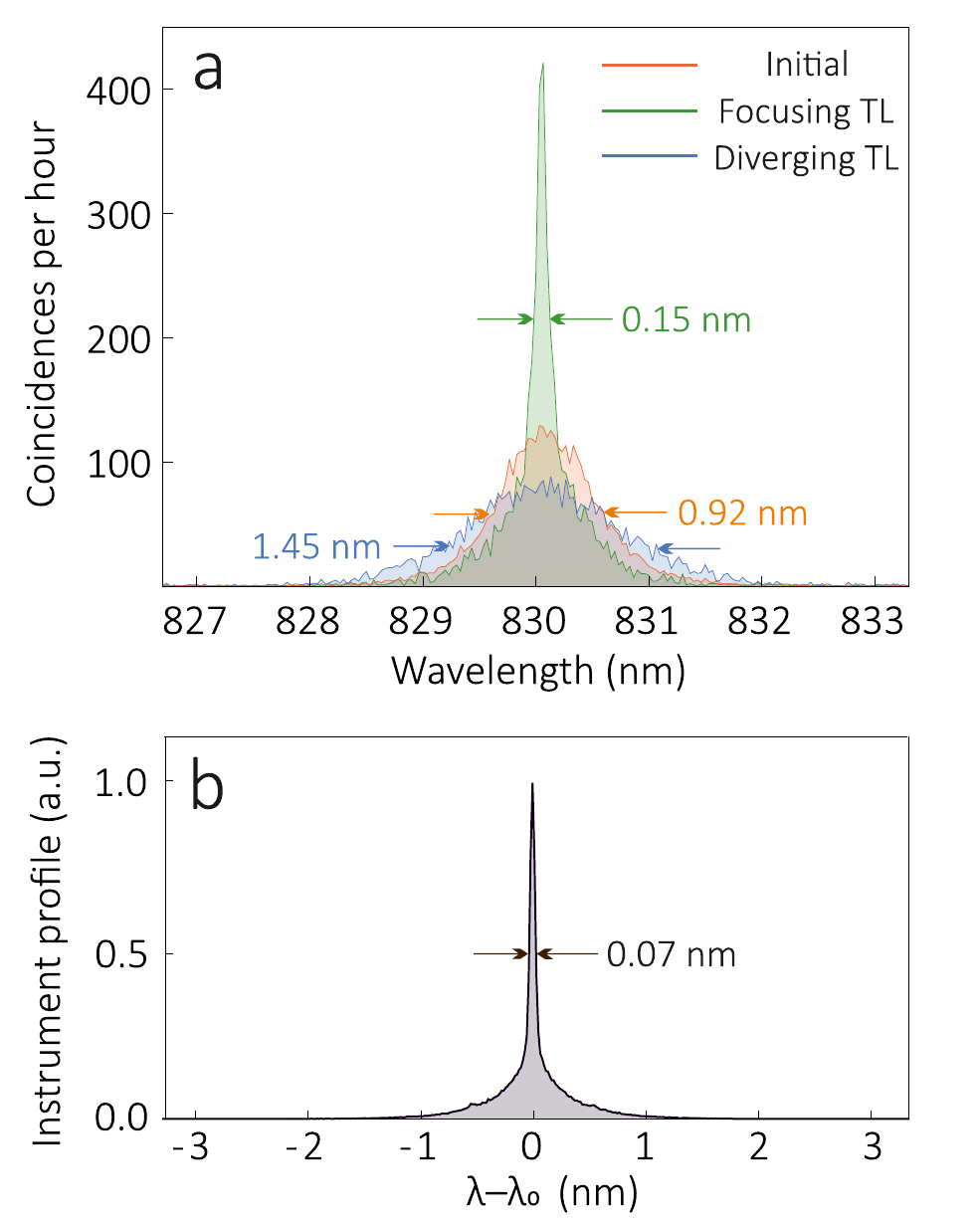}
\caption{\textbf{Spectral manipulation of single-photon wavepackets.} \textbf{a,} Raw spectrometer data presenting the initial (orange), compressed (green) and broadened (blue) spectra of heralded single-photon wavepackets corresponding to no time lens, focusing and diverging TL configurations, respectively. \textbf{b,} Instrument profile of the spectrometer employed in the experiment, which results in distortion of the measured spectral intensities.}
\label{Fig3}
\end{figure}

\begin{figure}[!th]
\includegraphics[width=0.88\columnwidth]{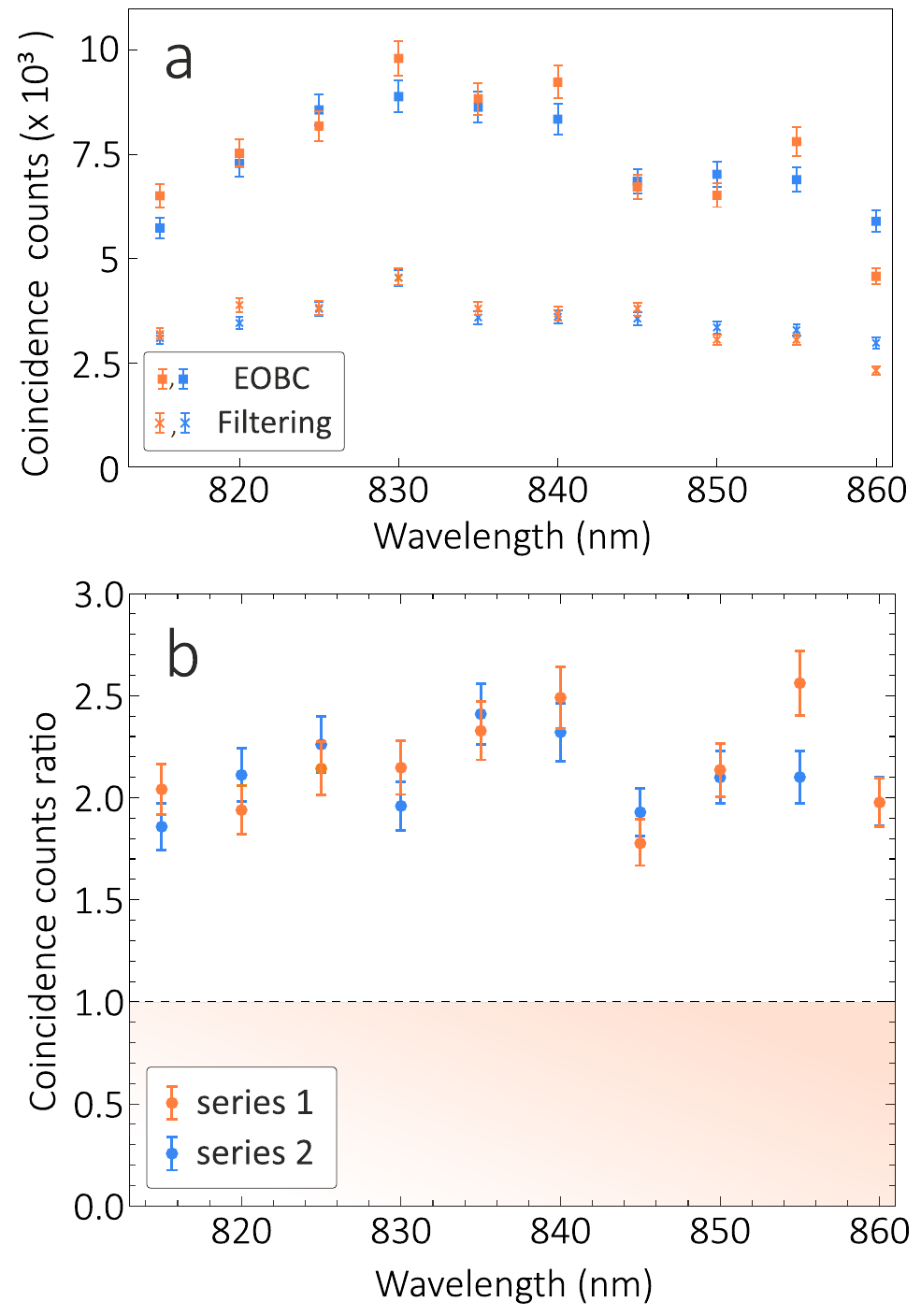}
\caption{\textbf{Performance of bandwidth compressor.} \textbf{a,} To verify the practicality of the EOBC we emulated spectrally-narrow single-photon absorption by measuring transmission through a narrowband spectral filter. Initially we directed the $3$~nm FWHM bandwidth heralded single-photon wavepackets onto a filter with $0.13$~nm to  $0.16$~nm FWHM spectral bandwidth, and detected the number of heralded photons transmitted through the filter in $150$~s (filled squares). Alternatively, we sent the heralded single-photon wavepackets through the EOBC before directing them through the filter and registering coincidence detection events (crosses). To demonstrate the versatility of this approach measurements were performed for a range of single-photon central wavelengths, with the filter tuned accordingly (see Methods for a detailed description of the experimental procedure). \textbf{b,} The ratio of counts registered in these two configurations remains above unity across the entire spectral investigated clearly confirming a net gain in photon flux to the narrowband absorber enabled by the spectral compression. Error bars combine contributions from Poisson count statistics and $\pm5\%$ relative count uncertainty originating from the non-repetitive transmission of fibre connectors used to switch between the two experimental configurations. This results in small discrepancies between the ratios obtained in two subsequent measurement series colour-coded with orange and blue.}
\end{figure}

We directly show that the EOBC can be used to enhance the rates at which bandwidth-incompatible photons can be successfully interfaced despite its non-unit overall transmission of $27\pm1\%$, (see Supplementary Information for discussion of device efficiency). To this end we emulated a spectrally narrowband absorber, such as a quantum memory, by a narrowband spectral filter (Methods). We measured the total flux of photons transmitted through a filter with transmission bandwidth comparable to the compressed photon bandwidth in two different experimental configurations: photons are (1) manipulated by the EOBC and then sent through the filter and (2) sent directly through the filter without passing through the EOBC, as presented in Fig.~2c. The numbers of coincidence counts registered in these two cases within an acquisition time of $150$~s are presented in Fig.~4a. In Fig.~4b ratios of coincidence counts measured in the two aforementioned scenarios are presented for a range of central wavelengths of the input photons (see Methods for details on source tuning and measurement procedure). The ratios significantly exceed unity across the entire $820$--$860$-nm spectral range, demonstrating the high efficiency and broad spectral acceptance of this approach to quantum pulse control. Indeed, this clearly demonstrates that our approach exceeds the performance of even unit efficiency spectral filtering. With these characteristics our technique can be readily applied in a sequence of spectral and temporal phases to achieve an arbitrary pulse-mode unitary transformation\cite{Morizur2010}, and holds promise for enabling efficient interfacing of hybrid quantum network nodes. The EOBC approach may also find application in spectral manipulation of general field-quadrature states of optical modes, such as squeezed states. Currently losses from the device introduce a significant admixture of vacuum noise to the quadrature state. However, it is feasible to address the technical sources of loss (see Supplementary Information for discussion of losses), thus improving the EOBC transmission to levels necessary for preservation of non-classical features of such states\cite{Vollmer2014}.

A key characteristic of a quantum pulse manipulation device is its performance in terms of the introduced photon noise, quantified by the conditional degree of second-order coherence, $g^{(2)}(0)$\cite{Zaske2012}. We experimentally determined $g^{(2)}(0)$ of heralded single photons with the EOBC switched on and off (Methods). The conditional degree of second-order coherence values without, $g^{(2)}(0)=0.0209\pm0.0024$, and with, $g^{(2)}(0)=0.0158\pm0.0031$, the EOBC switched on confirm preservation of the non-classical character of the heralded single-photon pulses, $g^{(2)}(0) < 1$. The use of RF fields to implement spectral control of optical pulses means that our device is intrinsically free from optical photon noise, which contrasts methods based on nonlinear optics, where care is needed to minimize contamination of the single-photon pulse with unwanted light originating from bright auxiliary optical pump beams.

We applied the technique of electro-optic temporal lensing to quantum light, experimentally demonstrating an all-fibre single-photon bandwidth compressor. We directly show that our device significantly increases the single-photon absorption rate into a spectrally narrowband absorber while preserving the non-classical character of single-photon wavepackets. The high-efficiency, broad spectral acceptance, reconfigurability of the applied phase, and central-wavelength preservation show the potential of electro-optic bandwidth manipulation for the development of quantum networks. To date electro-optic phase modulation methods have only been used to modify spectral-temporal properties of continuous-wave quantum light\cite{Olislager2014, Lukens2015}. Our work highlights the versatility of electro-optic manipulation as an efficient, low-noise platform enabling spectral control of both quantum and classical pulsed light sources over a broad range of central wavelengths. By utilizing both temporal and spectral phase operations this approach holds promise to advance TF-encoded quantum communication and simulation protocols\cite{Kielpinski2011, Brecht2015}, enabling tailored multimode TF unitary operations\cite{Morizur2010} by taking advantage of technology-driven developments in RF waveform generation\cite{Wang2015} and fibre Bragg gratings.

\footnotesize

\bibliographystyle{naturemag}



\newpage
\normalsize

\subsection*{Methods}

\footnotesize
\textbf{Single-photon source.} 
Pulses originating from a Ti:Sapphire oscillator ($80$~MHz repetition rate; Spectra-Physics Tsunami) with $200$~fs duration, $830$~nm central wavelength, are frequency doubled in a beta barium borate (BBO) crystal yielding a second harmonic beam with $415$~nm central wavelength and average power of $300$~mW. A small pick-off from the Ti:Sapphire beam is directed onto a fast photodiode (PD) to provide a synchronizing signal for the electro-optic time lens\cite{Kang2003}. The $415$~nm beam is focused in a $8$~mm long bulk potassium dihydrogen phosphate (KDP) crystal to probabilistically produce degenerate ($830$~nm central wavelength), spectrally-uncorrelated pairs of orthogonally polarized photons, signal and idler, with spectral bandwidths of $3$~nm and $8$~nm, respectively. Generated photons are separated using a polarising beam splitter (PBS) and coupled into single-mode fibres (Corning HI-780). By detecting the idler photon using a avalanche photodiode (APD) single photon counting module we herald the presence of the signal photon employed in the experiments. A field-programmable gate array (FPGA) is used to register coincidences between detection events with a coincidence window of $8$~ns for $g^{(2)}(0)$ and efficiency measurements, while a time-to-digital converter (TDC) is used to register coincidences for the heralded single-photon spectrum measurements.

\smallskip

\textbf{Single-photon spectrometer.}
The pulsed nature of photons used in the experiment enables us to employ a time-of-flight
spectrometer based on the dispersive Fourier transform principle. 
A chirped fibre Bragg grating (CFBG) maps the spectral
profile of a single-photon wavepacket onto its temporal distribution, which is
measured with respect to the idler photon detection\cite{Avenhaus2009, Davis2016}.
The temporal distribution is determined with resolution better than $50$~ps using
a pair of free-running low-timing-jitter single photon counting modules (Micro
Photon Devices, \$PD-050-CTB-FC) followed by a time-to-digital converter (TDC, Picoquant
PicoHarp 300) working in start-stop mode (see Fig.~2a). After appropriate
spectrometer calibration the time-of-arrival distribution
can be directly translated into the heralded photon spectrum. This approach
results in high spectral resolution of $0.07\mathrm{\:nm}$ with a spectral range of approximately $10$~nm, leading to a maximum temporal window of $10$~ns, which is compatible with the $80$~MHz pulse train. Finite spectrometer resolution results from timing jitter of the single photon counting modules monitoring the signal and idler arms. A more detailed description of the device and its calibration can be found in ref.\ \citen{Davis2016}.

\smallskip

\textbf{Spectral measurement uncertainty.} 
To estimate the uncertainties in spectral measurements we assume Poisson statistics for the experimentally collected photon counts in each spectral bin of width $0.034\mathrm{\:nm}$, with variance equal to the number of counts registered within this bin. Then we perform a Monte Carlo simulation by generating $10^{4}$ auxiliary spectra drawing the number of counts in each bin from the corresponding Poisson distribution, each time retrieving the FWHM bandwidth of generated spectra. This procedure produces $10^{4}$ estimates of the FWHM bandwidth related to one experimental spectrum and we take the standard deviation of this FWHM set as our measurement uncertainty. 

\smallskip

\textbf{Spectrally narrowband absorber.}
The narrowband absorber was realized as a bandpass spectral filter implemented
using a $300$-mm-focal-length monochromator (Andor Shamrock 303i) with a $1200$~lines/mm
diffraction grating. In place of the output slit at the focal plane
of the monochromator we placed the multimode fibre connected to an
avalanche photodiode single-photon counting module (Perkin Elmer SPCM-AQ4C). The grating was mounted on a motorized turret,
which allowed us to precisely tune the wavelength transmitted by the
filter. The filter transmission profile was measured using an optical
spectrum analyser (ANDO AQ6317) yielding spectral bandwidths of $0.13$ to $0.16$~nm FWHM, depending on the selected central wavelength.

\smallskip

\textbf{Determination of photon flux ratios.}
Experimental points in Fig.~4a are determined according to the following procedure.
We started by selecting the desired central wavelength of signal
photons by modifying the KDP crystal tilt angle. Afterwards we sent
photons directly into the narrowband absorber described in the previous
paragraph and maximized the registered coincidence counts rate by
rotating the monochromator grating. We collected the total coincidence
counts registered within $150$~s and present the obtained result in Fig.~4a (crosses).
 In the next step we inserted the EOBC between the single-photon source and
the narrowband absorber using two FC/PC-type fibre connectors. Using the manual phase-shifter of the RF signal we adjusted the
relative phase of the sinusoidal RF signal with respect to
the incoming photon wavepackets so that the coincidence count rate
is maximized. We compensated a small polarization rotation introduced
by bending the SMF during setup reconfiguration by manipulating a
polarization controller to maximize the coincidence count rate. We
collected the total coincidence counts registered within five series,
$30$~s each, monitoring the coincidence count rates and manually compensating
for RF signal phase drift if needed. The number of counts registered
in this configuration is presented in Fig.~4a (filled squares). 

\smallskip

\textbf{Second-order correlation function measurement.} 
The $g^{(2)}(0)$ measurement of heralded single photons was obtained by placing a polarization maintaining fibre coupler with a splitting ratio of approximately 50:50 in the signal arm with each output port monitored by an APD (Perkin Elmer SPCM-AQ4C), which we denote D2 and D3. Signals from D2, D3 and the heralding detector, denoted D1, were directed to the FPGA coincidence logic, where D1\&D2, D1\&D3 and D1\&D2\&D3 coincidence events as well as single count events at D1 were recorded within a coincidence window of $8$~ns. Counts were acquired in $50$ series of $360$~s duration. For each series the second-order correlation function $g^{(2)}(0)=\frac{N_{123}N_{1}}{N_{12}N_{13}}$ is calculated, where $N_d$ are counts within $360$~s for coincident detection events at a set of detectors labeled by $d$. We present the mean value of the calculated correlation functions with measurement uncertainty accounting for $5\%$ Ti:Sapphire laser power instability over the measurement time.

\footnotesize

\subsection*{Author contributions}

M.~K.\ and B.~J.~S.\ conceived the project. M.~K.\ and M.~J.\ designed and
performed the experiment. M.~J.\ analyzed the data with input from M.~K.\ 
L.~J.~W.\ developed the RF phase-locking system and contributed to
the early stages of the experiment. M.~K., M.~J., and B.~J.~S.\ wrote
the manuscript. M.~J.\ prepared the figures.

\subsection*{Acknowledgements} We acknowledge insightful comments and discussion about the work with K.\ Banaszek, C.\ Becher, A.~O.~C.\ Davis, E.\ Figueroa, D.\ Gauthier, X.\ Ma, C.\ Silberhorn, V.\ Torres-Company, N.\ Treps and I.~A.\ Walmsley. This project has received funding from the European Community (EC) Horizon 2020 research and innovation programme under Grant Agreement No.\ 665148. M.~K. was partially supported by a Marie Curie Intra-European Fellowship No. 301032 within the EC 7th Framework Programme and by the National Science Centre of Poland project No.\ 2014/15/D/ST2/02385. M.~J.\ was supported by the PhoQuS@UW project within the EC 7th Framework Programme (Grant Agreement No.\ 316244).

\subsection*{Competing Interests} The authors declare that they have no
competing financial interests.

\subsection*{Correspondence} Correspondence and requests for materials should be addressed to M.~J. (email: michal.jachura@fuw.edu.pl).

\normalsize

\clearpage
\onecolumngrid

\renewcommand*{\citenumfont}[1]{S#1}
\renewcommand*{\bibnumfmt}[1]{S#1.}

\makeatletter 
\renewcommand\@biblabel[1]{S#1.}
\makeatother

\setcounter{figure}{0} \makeatletter  \renewcommand{\thefigure}{S\arabic{figure}}

\section*{SUPPLEMENTARY INFORMATION}

\section{Choice of GDD to match electro-optic phase modulation depth}

To achieve bandwidth compression one needs to match the second-order spectral dispersion, also known as group delay dispersion (GDD), to the quadratic temporal phase. To determine the necessary GDD used in our bandwidth compression experiments we first retrieved the electro-optic phase modulation amplitude, $A$, using classical pulses transmitted through the electro-optic phase modulator under three different experimental configurations. We measured the central wavelength of the pulse spectrum when the RF signal is switched off and then measure the pulse spectrum with the RF signal switched on and the pulses aligned to the linear region of the RF signal resulting in a spectral shear up or down \cite{Duguay}. 

\begin{figure}[hbt!]
\includegraphics[width=0.5\columnwidth]{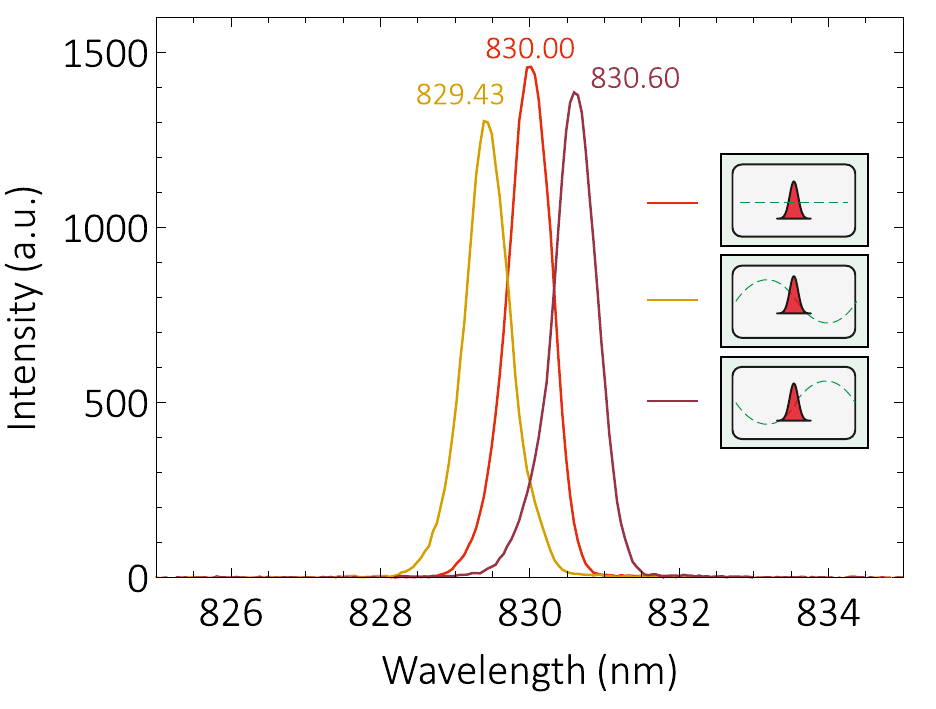}\centering
\caption{Spectrum of initial classical pulse (red, dashed)
phase-modulated by the means of EOM when its position is aligned
to linear region of the modulation within its positive (brown, dotted)
or negative slope (golden, solid). }
\end{figure}

The extremal spectral shifts correspond to the pulse aligned to either
positive or negative linear region of sinusoidal phase modulation
as presented in Fig.~S1. In these cases the sinusoidal phase modulation can be
approximated by the first-order linear term in series expansion: $\Delta\varphi(t)=A_{\mathrm{}}\sin(2\pi f_{\mathrm{RF}}t)\approx2\pi A_{\mathrm{}}f_{\mathrm{RF}}t.$
The time-dependent linear phase shift is analogous to the Doppler
effect imposing an angular frequency shift of $\delta \omega = 2\pi A_{\mathrm{}}f_{\mathrm{RF}}$
on the incoming light pulse. 

After measuring the central wavelength shift of $\Delta \lambda = 0.59 \pm 0.05$ nm corresponding
to the angular frequency shift of $\delta \omega =1.61 \pm 0.14 \mathrm{\times10^{12}\:rad}\mathrm{\,s}^{-1}$
we retrieved the phase modulation depth yielding $A = 25.7\pm 2.2$ rad, which results
in GDD of $\Phi=9.9 \pm 0.9 \mathrm{\:ps^{2}}$ necessary to meet the collimation condition for bandwidth compression.

\section{Modeling the spectral compression factor and absorption ratio}

Here we model the spectral compression factor and absorption ratio into our simulated narrowband absorber. First, we assume the initial pulses occupy normalized Gaussian spectral wavepackets with amplitude of the form
\begin{equation}
{\tilde{\psi}}(\omega) = \pi^{-1/4}\Delta \omega^{-1/2} \exp \left [  - \frac{(\omega - \omega_0)^2}{2 \Delta \omega^2} \right],
\end{equation}
centred on frequency $\nu_0 = \omega_0/2\pi$ with full-width at half-maximum (FWHM) intensity bandwidth $\Delta \nu_0 = \sqrt{\ln(2)} \Delta \omega / \pi$. Applying quadratic spectral phase $\phi(\omega) = \Phi (\omega - \omega_0)^2/2$, with the collimation condition $\Phi = 1/K$, followed by quadratic temporal phase $\theta(t) = Kt^2/2$, results in the following Gaussian spectral wavepacket amplitude
\begin{equation}
{\tilde{\psi}}(\omega) = \pi^{-1/4} \left(\frac{\Delta \omega}{i K}\right)^{1/2} \exp \left [  - \frac{i(K-i \Delta \omega^2) (\omega - \omega_0)^2}{2 K^2} \right].
\end{equation}
The spectral intensity of the output has the form
\begin{equation}
S(\omega) = |{\tilde{\psi}}(\omega)|^2 = \pi^{-1/2} \left(\frac{\Delta \omega}{K}\right) \exp \left [  - \frac{(\omega - \omega_0)^2}{(K/\Delta \omega)^2} \right],
\end{equation}
which yields FWHM intensity bandwidth
\begin{equation}
\Delta \nu' = \frac{\ln(2)}{\pi^2} \frac{K}{\Delta \nu_0}.
\end{equation}
The spectral bandwidth compression factor is given by the ratio of the input FWHM bandwidth, $\Delta \nu_0$ to the output FWHM bandwidth, $\Delta \nu'$,
\begin{equation}
C = \frac{\Delta \nu_0}{\Delta \nu'} = \frac{\pi^2}{\ln(2)}\frac{\Delta \nu_0^2}{K}.
\end{equation}
For our experiment at $830$~nm central wavelength, the input FWHM bandwidth is $\Delta \nu_0 = 401\pm26$~GHz and the temporal chirp parameter $K = 0.101 \pm 0.009$~ps$^{-2}$, which leads to a predicted compression factor of $23\pm4$, which is far from the value of $6.1 \pm 0.6$ observed in the experiment. This is due to fluctuations in the timing between the radio-frequency (RF) driving signal and the optical pulse train. A small temporal offset between the RF driving signal and the optical pulse train, $\tau$, results in an additional linear temporal phase term of the form, $\theta_1(t) = K \tau t$, which is equivalent to a spectral shift in the central frequency of the pulse. Assuming that the temporal offset is distributed by a Gaussian probability distribution centred about $\tau=0$ of the form $p(\tau) = \pi^{-1/2} T^{-1} \exp [-(\tau/T)^2]$, where $T$ is the width of the distribution, we can average over the timing offset, which leads to a broadened spectral output intensity
\begin{equation}
S'(\omega,T) = \pi^{-1/2} \left(\frac{\Delta \omega}{K}\right) \left(1+T^2\Delta \omega^2 \right)^{-1/2} \exp \left [  - \frac{(\omega - \omega_0)^2}{(K/\Delta \omega)^2(1+T^2\Delta \omega^2)} \right].
\end{equation}
This has a spectral FWHM intensity bandwidth
\begin{equation}
\Delta \nu'' = \frac{\ln(2)}{\pi^2} \frac{K}{\Delta \nu_0}\left(1+\frac{\pi^2}{\ln(2)}T^2\Delta \nu_0^2 \right)^{1/2},
\end{equation}
which results in a bandwidth of $70\pm7$~GHz with the temporal offset distribution width $T=0.5\pm0.05$~ps, which is consistent with the observed phase-locking stability in our experiment and the measured spectral bandwidth after the bandwidth compressor. The resulting predicted compression factor $C = 5.8\pm0.8$ matches the observed value of $6.1\pm0.6$ within the uncertainty values.

To determine the absorption ratio, we model the absorber by a Gaussian filter of width $\delta \omega_{\rm{F}}$ centred on the input pulse frequency $\omega_0$, with transmission of the form
\begin{equation}
g(\omega) = \pi^{-1/2} \delta \omega_{\rm{F}}^{-1} \exp \left [  - \frac{(\omega - \omega_0)^2}{\delta \omega_{\rm{F}}^2} \right].
\end{equation}
The flux of photons that are transmitted through the filter without use of the bandwidth compressor is given by the integral of the product of the filter transmission and the normalized input spectral intensity
\begin{equation}
F_0 = \int_{-\infty}^{\infty}g(\omega)|{\tilde{\psi}}(\omega)|^2 {\rm{d}}\omega,
\label{f0}
\end{equation}
which results in a fraction of transmitted photons without the bandwidth compressor given by $F_0 = [1+ (\Delta \omega /\delta \omega_{\rm{F}})^2]^{-1/2}$. By replacing the input spectral intensity with the output spectral from the bandwidth compressor in Eq. (\ref{f0}), and multiplying by the transmission through the bandwidth compressor, $\eta$, yields the fraction of photons transmitted through the filter after passing through the bandwidth compression, given by $F_1 = \eta  [1+ (K / \Delta \omega \delta \omega_{\rm{F}})^2(1+T^2\Delta \omega^2)]^{-1/2}$. Thus, the absorption ratio is given by the ratio of the fraction of photons transmitted when using the bandwidth compressor divided and the fraction of photons transmitted without using the bandwidth compressor
\begin{equation}
R = \frac{F_1}{F_0} = \eta \left[\frac{1+\left(\Delta \omega / \delta \omega_{\rm{F}} \right)^2}{1 + \left(K/\Delta \omega \delta \omega_{\rm{F}} \right)^2 \left( 1+T^2\Delta \omega^2 \right)} \right]^{1/2}.
\end{equation}
With a spectral filter with FWHM intensity bandwidth $\delta \nu_{\rm{F}} =  \sqrt{\ln(2)} \Delta \omega / \pi = 57\pm5$~GHz as used in the experiment, along with the transmission through the bandwidth compressor of $\eta =0.27 \pm 0.01$ and same values for the timing offset distribution width and initial pulse bandwidth we predict the absorption ratio $R = 1.8\pm0.2$, which matches the measured value of $2.0 \pm 0.2$ within the experimental uncertainties.

\section{Spectral compression with $40$~GHz time lens}

Maximum compression of spectral bandwidth using the electro-optic bandwidth compressor (EOBC), comprised of spectral dispersion and a time lens (TL) implemented by quadratic temporal electro-optic phase modulation, is achieved when the entire temporal aperture of the TL is used (assuming a transform limited pulse enters the dispersive element). When this condition is satisfied the spectral bandwidth of the field
exiting the EOBC is inversely proportional to the TL temporal aperture, which is set by the RF sinusoidal modulation frequency $f_{\mathrm{RF}}$.
The optimal input field spectral bandwidth is determined by the chirping
factor $K$, arising from the quadratic temporal phase of the TL, $\theta(t) = Kt^2/2$, or equivalently (due to the collimation condition $\Phi = 1/K$)
to the GDD applied to the pulse, $\Phi$, and corresponds to the temporal duration
of the chirped pulse completely filling the TL aperture. For a given
modulation frequency, $f_{\rm{RF}}$, the chirp parameter, $K = 4 \pi^2 f_{\rm{RF}}^2 A$, is proportional to the modulation depth $A$. Thus input and output bandwidths can be independently tuned by modifying the modulation frequency and depth (provided they stay within experimentally realistic parameters). Here we show that the same procedure as that used for the $10$~GHz TL, can be used to implement three-fold spectral compression of $1.98 \pm 0.08$~nm bandwidth heralded single-photon pulses using a $40$~GHz modulation frequency. The modulation depth for our $40$~GHz modulator (EOSpace PM-AV5-40-
PFU-PFU-830-S) was found to be $A = 7.2$~rad, resulting in a GDD of $2.2\;\mathrm{ps^{2}}$, required to achieve the collimation condition, which we introduced by means of $56.5$~m of single-mode fibre (Corning HI-780). A spectrally compressed (expanded) wavepacket of $0.66 \pm 0.04$~nm ($3.2 \pm 0.3~\text{nm}$) bandwidth was obtained. Results of spectral intensity measurements for this experiment are presented in Fig.~S2. 

\begin{figure}[hbt!]
\includegraphics[width=0.5\columnwidth]{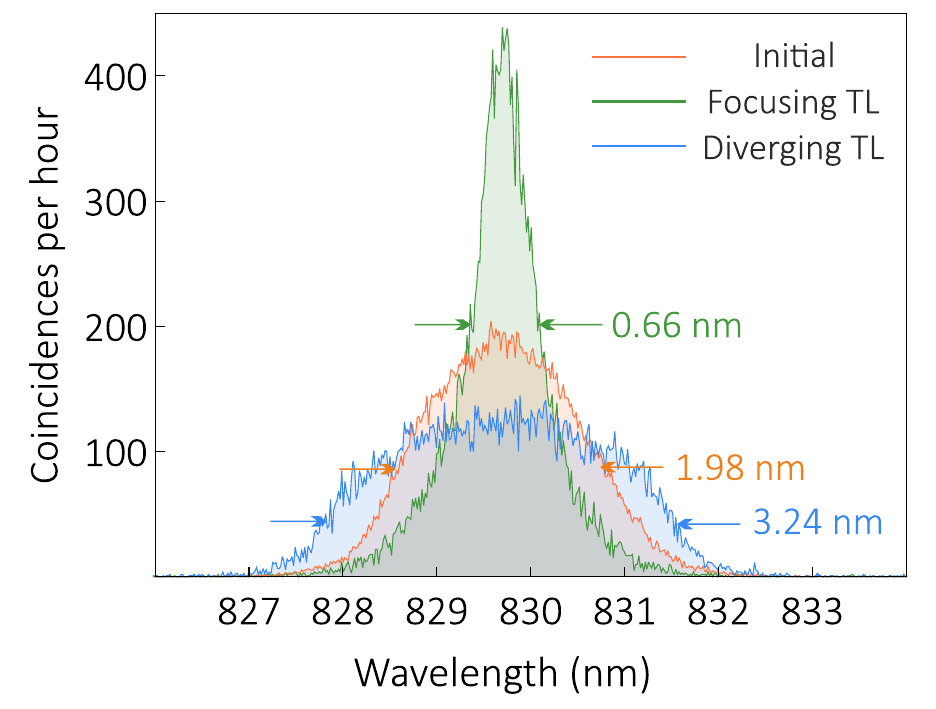}\centering

\caption{Spectral compression (expansion) of single-photon wavepackets obtained using $40$~GHz electro-optic modulation frequency. Raw spectrometer data present both the initial (orange) and compressed (green) spectrum of the heralded single-photon wavepackets. To illustrate the space-time analogy between electro-optic bandwidth compression and Gaussian beam collimation we also collected the spectrum of heralded photons transmitted through a diverging rather than focusing time lens (blue).}
\end{figure}

\section{Electro-optic bandwidth converter efficiency}

The overall efficiency of a bandwidth converter can be expressed as the product of the intrinsic conversion efficiency of the bandwidth conversion process and the transmittance through the device. Transmission is primarily limited to less than unity by technical losses and does not pose a fundamental obstacle. However, achieving unit conversion efficiency -- and thus deterministic operation -- is challenging to realize for nonlinear optical processes such as three- or four-wave-mixing. In contrast the electro-optic phase modulation method introduced here is intrinsically deterministic -- any pulse passing through the modulator is modified. 

The conversion efficiency of the phase-modulation process is unity. This is due to the fact that the application of radio-frequency electric fields, such as those considered here ($10$~GHz and $40$~GHz), to the electro-optic medium results in deterministic modification of the refractive index of the material through the Pockels effect. The unit conversion efficiency of electro-optic phase modulation contrasts the efficiency of nonlinear parametric bandwidth conversion methods, based on three- and four-wave mixing, which generally operate in a probabilistic manner in which not every wavepacket passing through the medium is modified. Conversion efficiencies approaching $80\%$ have been demonstrated for single-photon frequency conversion by means of three-wave \cite{Vollmer2014, Ates2012, Zaske2012} and four-wave mixing \cite{Clemmen2016}. However comparable efficiencies for bandwidth conversion of single-photon wavepackets have not yet been demonstrated. Even though spectral bandwidth conversion efficiency of $40\%$ by four-wave mixing has been shown for classical weak coherent pulses \cite{Agha2014}, for non-classical light the highest reported internal conversion efficiency was up to $1.1\%$ \cite{Fisher2016}.  Nevertheless, recent theoretical works show that unit conversion efficiency for parametric frequency conversion might be in general impossible for pulsed single-photon wavepackets due to Hamiltonian time-ordering effects \cite{Christ2013, Quesada2016}. There are strong reasons to believe that these limitations equally apply to bandwidth conversion. No such limitations are present in direct phase modulation approaches \cite{Matsuda2016}, including our electro-optic scheme.

Whereas the internal conversion efficiency of the electro-optic time lens can be considered unity, the overall transmission of the $10$~GHz electro-optic bandwidth compressor (EOBC) is reduced to $27\%$ due to several technical sources of loss, namely (1) propagation losses in the dispersive fibre, (2) electro-optic modulator losses, and (3) losses at fibre cable interfaces.

To introduce the required amount of group delay dispersion (GDD) to match the $10$~GHz time lens $256$~m of HI-780 single-mode fibre (Corning) was used. The fibre cable was connectorized with FC/PC connectors at both ends. Transmission through $256$~m of HI-780, with a loss figure of $3.4$~dB/km as per manufacturer specification, equals $0.82$. Losses associated with the connectors will be discussed later. For the amount of GDD required here a length of optical fibre is currently the optimal choice in terms of losses. For higher amounts of GDD chirped fibre Bragg gratings may become advantageous. Note that any bandwidth compression device requires GDD and will thus incur similar losses.

The electro-optic phase modulator (EOSpace PM-AVe-40-PFU-PFU-830) comprised an $8$~cm long lithium-niobate waveguide pigtailed with $1$~m of Panda-style polarization-maintaining  (PM) fibres at both ends. The fibre pigtails were terminated with FC/PC connectors. We measured the transmission through the pigtailed modulator at $0.39 \pm 0.04$, excluding losses at FC/PC connectors. The principal source of loss is transverse-spatial mode mismatch between the fibre pigtails and the waveguide in lithium niobate modulator \cite{Moyer1998}, which could be reduced by appropriate tapered structures. Alternatively the losses could be reduced by using a bulk electro-optic modulator at the cost of significantly increased half-wave voltage.

The final contribution to EOBC losses comes from fibre connectors. The dispersive fibre was connected to the EOM pigtail by an FC/PC--FC/PC connector interface. Whereas the dispersive fibre was a single-mode HI-780, the EOM fibre pigtail was a $5~\mu$m core diameter Panda-style polarization-maintaining (PM) fibre. Additionally, FC/PC connectors were used to feed the optical signal into and out of the EOBC. Thus, inserting the EOBC into a fibre-optic link introduces two additional FC/PC--FC/PC connector interfaces. Typical loss of an FC/PC--FC/PC interface is $0.3$~dB, yielding $13\%$ loss for the two connector interfaces. A small additional loss contribution might arise due to mode mismatch between the HI-780 and the Panda-PM fibre. The connector losses could be significantly reduced by replacing fibre connector interfaces with fibre splices and by using an HI-780-pigtailed EOM.

The combined estimated EOBC transmission due to the three contributions listed above is $0.28\pm0.04$. The estimated value agrees well with the experimentally determined EOBC transmission of $0.27\pm0.01$. The transmission of the EOBC was determined as follows: attenuated pulses from the Ti:Sapphire oscillator were launched into an FC/PC-connectorized PM-fibre patchcord followed by a PM-fibre-pigtailed optical power meter. After recording the reference power meter reading $P_0$, the EOBC was inserted between the patchcord and the power meter pigtail. Power meter reading $P_1$ was recorded and the ratio $T = P_1/P_0$ calculated. The procedure was repeated $5$ times; each iteration included reconnecting the FC/PC--FC/PC interface between the modulator and the dispersive fibre. The reported EOBC transmission is the mean of the obtained $T$ values, with measurement uncertainty determined according to the $t$-Student distribution. The EOM transmission estimate is based on a single measurement, with uncertainty corresponding to the Ti:Sapphire laser intensity variations. The $27\%$ device efficiency of our EOBC exceeds the device efficiencies of bandwidth compression experiments on quantum light currently reported by factors exceeding $4 \times10^2$ and $30$ \cite{Lavoie2013,Fisher2016}.

A possible future application of the EOBC approach could be in faithful spectral manipulation of continuous-variable field-quadrature states of an optical mode. Such manipulation could be interesting to enable monitoring large-bandwidth squeezing with bandwidth-limited balanced homodyne detectors \cite{Patera2015}. Current EOBC device efficiency of $27\%$ is not high enough to enable useful manipulation of such fragile states, since on every passage through the device the state of the optical mode would be mixed with a $73\%$ vacuum contribution, detrimental to non-classicality of interesting continuous-variable states of an optical mode, such as squeezed or cat  states. Since transmission of EOBC is limited only by technical losses, we see it feasible for the EOBC device efficiency to approach, and possibly exceed, the $50\%$ device efficiency level which enables preservation of significant amounts of squeezing \cite{Vollmer2014}. The necessary enhancement can be achieved by improving spatial mode matching within the EOBC.

\section{Electro-optic bandwidth converter noise}

Noise in spectral-temporal mode converters can by divided into noise that affects the electric field quadratures and noise in the modes into which light is converted. The most common field-quadrature noise is caused by the introduction of additional photons during the mode conversion process, and results in modified photon number statistics at the output. Owing to the complete lack of auxiliary optical fields our electro-optic scheme is free from such field-quadrature noise, which we experimentally verify by measuring the second-order correlation function value. This is in contrast to both parametric \cite{Lavoie2013, Vollmer2014, Ates2012, Zaske2012, Fisher2016, Clemmen2016} and non-parametric \cite{Matsuda2016} nonlinear optical approaches, where special care needs to be taken to avoid contamination of the single-photon signal with photons scattered from strong pump beams. 

Noise associated with the mode distribution into which light is converted in our setup comes from drift in the temporal offset between the optical pulse train passing through the electro-optic modulator and the radio-frequency (RF) field driving the modulator. This temporal offset determines the temporal phase experienced by the optical pulse. The drift happens on two time scales: timing jitter on a time scale of approximately $0.5$~ps and slow drift on a time scale of approximately $1$~minute and manifests itself in incoherent spectral broadening of the output pulse, as we show in the {\em Modeling the spectral compression factor and absorption ratio} supplementary section. We compensated for the slow drift in our experiment by acquiring data in intervals of up to $60$~s duration, followed by recalibrating the phase of the RF field. We experience spectral broadening due to the timing jitter and residual slow drift effects. The slow drift could be significantly reduced by temperature stabilizing the electro-optic modulator, possibly combined with an active feedback loop based on spectral measurements of a wavelength-multiplexed bright auxiliary beam. The timing jitter could be reduced to a negligible level by using a more advanced optical-RF synchronizing scheme, such as in ref.\ \citen{Jung2012}. 

A natural extension of this work would be direct experimental verification of coherence across the spectral modes of the bandwidth compressed pulse. Although some degree of coherence between spectral modes is implicitly verified by virtue of the frequency-to-time mapping performed by the chirped fibre Bragg grating spectrometer used to characterize the output light, this does not directly verify the degree of coherence. A possible method to perform such verification would be via complete quantum state tomography in the time-frequency domain \cite{Wasilewski2007,Ansari2016}, although several experimental challenges would need to be addressed to adapt these methods to match the narrow bandwidth of spectrally compressed pulses. A more direct method could involve two-photon, or Hong-Ou-Mandel, interference with a pure reference single-photon wavepacket with spectral profile, including spectral phase, matched to the bandwidth-compressed pulse. Such reference pulse could be prepared either by appropriately engineering a heralded single photon source, or by spectral filtering  with an etalon from broadband single photons wavepackets, possibly combined with pulse shaping by means of highly dispersive media for spectral phase compensation. High-visibility Hong-Ou-Mandel interference with a matched reference pulse would confirm the purity of the spectrally manipulated time-frequency state of the single photon wavepacket \cite{Hendrych2003, Wright2016}.

\end{document}